\documentclass[12pt,a4paper]{article}
\usepackage{a4wide}

\begin{document}

\begin{center}

{\Large \bf Localizable Effective Theories, Bootstrap
and the Parameters of Hadron Resonances} \\
\vspace*{0.5cm}

K.~Semenov-Tian-Shanski$^a$,
\underline{A.~Vereshagin}$^{a,b}$ and
V.~Vereshagin$^a$ \\
$^a$St.Petersburg State University, 
$^b$University of Bergen
\vspace*{0.5cm}

{\em Talk given at the X. International Conference On
Hadron Spectroscopy (HADRON'03), \\
August 31 - September 6, 2003,
Aschaffenburg, Germany}

\end{center}

\begin{abstract}
We discuss the basic principles of constructing a meaningful
perturbative scheme for effective theory. The main goal of this talk
is to explain the approach and to present recent results
\cite{VV} -- \cite{AVVV3} obtained with the help of the method of Cauchy
forms in several complex variables.
\end{abstract}

Our work is aimed to develop a field-theoretic scheme providing the
basis for dual models and, simultaneously, a link between two methods
-- the quantum field theory and the analytic theory of $S$-matrix.
Besides, it concerns with the problem of renormalization of
conventionally nonrenormalizable theories.

We rely on Weinberg's general scheme (see Chs. 2-5 of \cite{WeinQFT})
of constructing a quantum theory in terms of field operators that
create and annihilate the {\em true} asymptotic states (corresponding
to stable particles). Following \cite{WeinEFT}, we call a theory
effective if the interaction Hamiltonian (in the interaction picture)
contains {\em all} terms consistent with the requirements of a given
algebraic (linear) symmetry. The scheme is quantum {\em ab initio} and
the problem of dynamical (nonlinear) symmetries requires special
consideration. Here we do not discuss it.

We only consider
 a special class of
{\em localizable}
effective theories. An
\underline{initial}
effective theory (constructed  according to Weinberg's scheme solely
from the fields of true asymptotic states) is called localizable if
its tree level amplitudes can be reproduced in the framework of the
well-defined
\underline{tree level}
approximation based on the Hamiltonian of
\underline{extended}
effective theory which -- along with the fields of stable particles --
also contains complementary auxiliary fields corresponding to
fictitious unstable particles (resonances) of arbitrary high spin and
mass. This implies that infinite sums of graphs providing formal
expressions for tree level amplitudes of the initial theory converge,
at least in certain small domains of the corresponding complex spaces.
The existence of such domains is, in any case, necessary to assign
meaning to the initial theory. It is precisely those domains where the
tree level amplitudes of both theories must coincide with each other.
Thus the extended theory just provides an analytic continuation of the
tree level amplitudes constructed in the framework of the  initial
effective theory. In fact, our approach is just an attempt to extend
Weinberg's quasiparticle method (see
\cite{WeinQuasi}
and refs. therein) to the case of relativistic quantum theory.

It should be stressed that we only consider the case when the extended
effective theory (as well as the initial one) does not contain
massless particles of spin J > 1/2. Besides, we are only interested in
constructing the effective
{\em scattering}
theory -- the calculation of Green functions is not implied. This
means that the renormalization procedure possesses certain specific
features allowing one to avoid attracting unnecessary renormalization
prescriptions. The divergences which might occur in Green functions
off the mass shell never bother us.

The last point deserves comment. In the case of customary
renormalizable theories the number of renormalization prescriptions
(RP) fixing the finite parts of counterterms is equal to that of
coupling constants in the Hamiltonian (including mass and kinetic
terms),
{\em this latter one being finite}.
``Hidden''
couplings create no problem (see, e.g.,
\cite{Collins}).
In case of effective theories the situation looks quite different.
They are renormalizable by construction because all possible local
monomials are presented in the Hamiltonian. The problem is that the
number of coupling constants is essentially infinite. This means that,
to obtain finite results for all Green functions, one needs to point
out a complete self-consistent infinite set of independent RP's. The
structure of this set must provide a guarantee of convergence of
infinite series appearing at every given order of loop expansion due
to the presence of field derivatives of arbitrary high degree and
order in the effective Hamiltonian. Otherwise, the resulting
amplitudes would make no sense. The problem of constructing of such a
set looks unsolvable until we have no regularity fixing its structure
(possibly, up to a finite number of independent constants). The
requirement of localizability is extremely useful in this very
respect. As shown in
\cite{AVVVKS}, \cite{AVVV3},
in case of effective scattering theory it is possible to perform a
detailed classification of parameters (combinations of coupling
constants) appearing in expressions for the renormalized
$S$-matrix
elements of extended effective theory. This allows one to separate a
group of
{\em resultant}
parameters. This group only contains the parameters which do
contribute to renormalized $S$-matrix elements and thus require
formulating the RP's. Finally, it is possible to show that crossing
symmetry together with the requirement of convergence impose strong
limitations on  the allowed values of resultant parameters. In the
case of tree level parameters describing  the amplitudes of binary
processes these limitations take a form of an infinite system of
(algebraic)
{\em bootstrap}
equations connecting the values of the resultant coupling constants
and the mass parameters appearing in the Hamiltonian. It is a direct
consequence of the postulated properties of meromorphy and polynomial
boundedness of the tree level amplitudes in every 3-dimensional band
$B_x\{x \in {\bf R},\ x \sim 0;\ {\nu}_x \in {\bf C}\}$,
where
$x$
stands for (real) momentum transfer and
${\nu}_x$ --
for the corresponding (complex) energy-like variable.

We demand such properties for the following reasons. First, the
polynomial boundedness property of the full (non-perturbative)
amplitudes follows from the general axiomatic requirements. Hence it
makes sense to construct the perturbation series in such a way that at
every step the perturbative amplitude  possesses the property of
polynomial boundedness with the same degree of bounding polynomial as
that of the full (non-perturbative) amplitude -- this gives a chance
to avoid strong corrections from the higher order terms. Besides, this
requirement provides a guarantee that we deal with tempered
distributions. Second, the property of meromorphy of tree level
amplitudes follows from the
{\em summability}
requirement. The meaning of this term can be explained as follows. At
every given order of loop expansion we have to take account of an
infinite number of graphs. In absence of guiding principle fixing the
summation order there is a danger either to get diverging series or to
fall in contradiction with basic properties like crossing symmetry.
That is why it looks reasonable to impose the summability requirement
in the following form.
{\em In every sufficiently small domain of the complex space of
kinematical variables there must exist an appropriate order of
summation of the formal sum of contributions coming from the graphs of
a given loop order, such that the reorganized series happens
convergent. Altogether, these series must define a unique analytic
function with only those singularities which are presented in
contributions of individual graphs or can be reproduced in the
framework of the same loop order of extended theory containing
auxiliary fields corresponding to unstable particles.} %
This definition can be considered as a generalization (or, better,
detailing) of the localizability requirement. From this formulation it
follows that to study the most general localizable effective theory it
is quite sufficient to consider the case of extended theory with no
{\em ad hoc}
limitations on the structure of set of resonance parameters. In turn,
this means that the tree level amplitudes of the extended theory must
be meromorphic functions of kinematical variables. They must be
polynomially bounded in certain energy-like variables at fixed values
of the other ones,the degree of bounding polynomials being dictated by
unitarity.

Here it is pertinent to stress one point. To present the results of
resonance saturation of dispersion relations (DR) for the amplitudes
with nondecreasing asymptotic behavior the following form is often
used:
\begin{equation}
A(z,x) = P(z,x) + \sum\limits_{k=1}^{\infty}
\left[ \frac{r_k(x)}{z - p_k(x)} \right].
\label{1}
\end{equation}
Here
$z$
stands for energy-like variable, $x$ -- for any real parameter (say,
the momentum transfer) and $P(z,x)$ -- for so-called subtracting
polynomial in
$z$
with the coefficients depending on
$x$.
However, it should be noted that in the case of infinite number of
resonances this form is to be taken just as a formal one. Usually, it
is tacitly assumed that the series of pole contributions in the RHS
converges. Unfortunately, this is not always true
(see, e.g., \cite{Shabat}).
This suggestion imposes too strong limitations on the values of
resonance parameters
$r_k(x)$
and
$p_k(x)$.
In fact, it states that resonances do not affect the asymptotic
behavior of the amplitude in question. In case of strong interactions
this looks too restrictive. For this reason in our work we use the
method of Cauchy forms specially adjusted for the case of several
variables (see
\cite{AVVV2}, \cite{AVVVKS});
this allows us to avoid implicit postulating of the resonance spectrum
properties. Using this method it is easy to write down the most
general form of the result. In case when the amplitude  has only
simple poles at
$z_k = p_k(x) \neq 0$
with residues
$r_k(x)$
and behaves, say, like a constant (it only makes sense to speak about
the contour asymptotics) for
$x \in (a,b)$
and
$|z| \to \infty$,
it looks as follows
\begin{equation}
A(z,x) = A(0,x) + \sum\limits_{k=1}^{\infty}
\left[
\frac{r_k(x)}{z - p_k(x)} + \frac{r_k(x)}{p_k(x)}
\right],\ \ \ \ \ \ \ \
x \in (a,b).
\label{2}
\end{equation}
It is possible to show (see
\cite{AVVV2})
that in this case the functional series
$
S_n = \sum_{k=1}^{\infty} \frac{r_k(x)}{p_k^n(x)}
$
is certainly convergent for
$n \geq 2$
and may happen divergent for
$n< 2$.
In case when
$S_1 \leq M < \infty \;$
the only possibility to fulfil the postulated asymptotic behavior is
to demand that
$A(0,x)\neq 0. \;$
This means that the asymptotics is completely formed by
$A(0,x)$,
i.e., the contribution of resonances is irrelevant. These problems
disappear if one uses the Cauchy form
(\ref{2}).

The summability requirement imposes certain restrictions on the
parameters of theory. As shown in cited above papers, those
restrictions for the tree level parameters follow from the condition
of identical coincidence of two Cauchy forms representing the
amplitudes of cross-conjugated processes. Each one of these forms is
only applicable in the corresponding 3-dimensional band (layer). In
the intersection of those layers both forms are equally applicable and
thus must coincide. This requirement leads to an infinite system of
algebraic relations between the resultant parameters.

The corresponding mechanism may be illustrated by the following
example. Consider a rational function of two complex variables
$F(x,y)$.
Let us suppose it has a single simple pole (in
$x$)
in the layer
$
B_y \{ \ y  \in {\bf R},\  y \in (-\eta , +\eta);\ x \in {\bf C} \}
$
and also a single simple pole (in
$y$)
in the `orthogonal' layer
$
B_x \{ \ x \in {\bf R},\  x \in (-\xi, +\xi);\  y \in {\bf C} \}.
$
Let us also assume that
$F(x,y)$
is decreasing at infinity in each layer and that it is regular at the
origin
$M(0,0)$.
Now let us try to answer the following question: what is the
structure of the set of numerical parameters providing a complete
description of functions that possess these properties?

Every function regular at the origin is completely fixed by the
coefficients
$f_{ij}$
of its series expansion
$F(x,y) = \sum_{}^{} f_{ij} x^i y^j$.
The above question can be rephrased in a more concrete way: how
many independent combinations of these coefficients can be % fixed
arbitrary and what are these combinations?  Or, in  terms of field
theory: how many independent renormalization prescriptions is it
necessary to impose in order to completely fix the amplitude
$F(x,y)$
and what is the explicit form of those prescriptions?

In the layer
$B_y$
we have
$
F(x,y)  =
\frac{\rho (y)}{x - \pi (y)}.%\ \ , \ \ \ \ \ \ \ \ (x,y) \in B_y\ .
$
By condition, the functions
$\rho (y)$
and
$\pi (y)$
are regular in the vicinity of the origin. Hence,
$\pi (y)  = \sum \limits _{}^{} {\pi}_i y^i$, % \ \ \ \ \ \ \
$\rho (y) = \sum \limits _{i=0}^{} {\rho}_i y^i$.
By analogy, in
$B_x$:
$
F(x,y)  =
\frac{r(x)}{y - p(x)} , %\ \ \ \ \ \ \ \ (x,y) \in B_x\ ,
$
where
$p(x) = \sum_{}^{} p_i x^i$ ,% \ \ \ \ \ \ \
$r(x) = \sum_{}^{} r_i x^i$.
Hence in the intersection domain
$ B_x \cap B_y \equiv D_{xy} $:
\begin{equation}
\frac{r(x)}{y - p(x)} = \frac{\rho (y)}{x - \pi (y)}\; ,
\ \ \ \ \
(x,y) \in D_{xy}
\{x \in (-\xi, +\xi) ,\; y \in (-\eta , +\eta)\} \ .
\label{2.8}
\end{equation}
Substituting
$\pi(y)$,
$\rho(y)$,
$p(x)$
and
$r(x)$
into
(\ref{2.8}),
we obtain an infinite system of conditions on the coefficients
$p_k, r_k, {\pi}_k, {\rho}_k$:
\begin{equation}
r_{i+1} {\pi}_0 - p_{i+1} {\rho}_0 = r_i, \ \ \ \
{\rho}_{i+1} p_0 - {\pi}_{i+1} r_0 = {\rho}_i, \ \ \ \
r_{i+1} p_{j+1} = {\rho}_{i+1} {\pi}_{j+1}\ \ \ \ \ i,j=0,1,...
\label{2.9}
\end{equation}
This system provides an example of what we call the bootstrap
equations. Once solved, it permits to express the parameters
$p_i, r_i$
in terms of
${\pi}_i, {\rho}_i$.
So it gives an answer to the question wether it is possible to carry
out the analytic continuation from one layer to another. This is an
infinite system of equations with respect to
$2 \times \infty$
(formal notation!) unknown parameters, needed to reexpress the
function
$F(x,y)$
in the layer
$B_x$
in terms of the parameters  defining it in the layer
$B_y$.
In general, it is very difficult to find solutions of such systems and
even to show their solvability. Fortunately, in this simple example it
turns out possible to write down the solution in explicit form.
Separating the variables in
(\ref{2.8}),
taking derivatives and solving the corresponding ordinary differential
equations, one finds:
\begin{equation}
F(x,y) = \frac{ad+bc}{-d + axy + bx + cy}.\
\label{2.14}
\end{equation}
The important property of this result is that it contains only 4
arbitrary parameters! This means that the infinite system
(\ref{2.9})
only happens consistent if the function
$F(x,y)$
defined in the layer
$B_y$
belongs to the four-parametric family (\ref{2.14}). This is the only
case when there exists the analytic continuation of this function from
$B_y$
into
$B_x$
with the desired properties. It is clear that in this case the
continuation is unique.

This exercise gives an idea of the
``power''
of bootstrap restrictions. The direct analysis of the system
(\ref{2.9})
would lead to the same conclusion. In this simple example it happens
possible. Unfortunately, the regular method of solving infinite
algebraic systems is not known, except few trivial cases.

With the help of
(\ref{2.14}),
one can express the parameters
$
f_{ij} = f_{ij}(a_1,a_2,a_3,a_4)
$
in terms of ``fundamental constants''
$a_i$ $(i=1,\ldots ,4)$.
Then one can choose four arbitrary coefficients
$f_k$ ($k=1,2,3,4$)
(or four arbitrary combinations) that allow the inversion
$
a_i=a_i(f_1,...,f_4),
$
and impose arbitrary ``renormalization prescriptions'' for these four
quantities. The values of all other parameters should respect the
conditions
(\ref{2.9}).

So,
{\em to fix the amplitude}
$F(x,y)$
{\em uniquely }
{\em it is sufficient to impose four independent renormalization
prescriptions defining the
``fundamental''
constants}
$a,b,c,d$.

Precisely the same mechanism provides the system of bootstrap
constraints for the parameters of pion-nucleon resonances (see
\cite{piNAVV}).
The most important feature of this system is the renormalization
invariance:
{\em bootstrap equations are nothing but the restrictions for
renormalization prescriptions}. %
This very property allows us to compare the bootstrap equations
directly with known experimental data.

\subsection*{Acknowledgments}

The work was supported in part by INTAS (project 587, 2000), RFBR
(grant 01-02-17152) and by Russian Ministry of Education
(Programme Universities of Russia, project 02.01.001). The work by
A.~Vereshagin was supported by Meltzers H\o yskolefond
(Studentprosjektstipend 2003).


\begin{thebibliography}{99}
\bibitem{VV}
 V.~Vereshagin, 
 {\em Phys. Rev. D} {\bf 55}, 5349 (1997).
\bibitem{AVVV2}
 A.~Vereshagin and V.~Vereshagin,
 {\em Phys. Rev. D} {\bf 59}, 016002 (2000).
\bibitem{AV}
 A.~Vereshagin,
 {\em $\pi N$ Newsletter} {\bf 16}, 426 (2002).
\bibitem{AVVVKS}
 A.~Vereshagin, V.~Vereshagin, and K.~Semenov-Tian-Shanski, 
 {\em Zap. Nauchn. Sem. POMI} {\bf 291}, Part 17, 78 (2002); English    
 version is to appear in
 {\em J. Math. Sci. (NY)}, (2003).
\bibitem{AVVV3}
 A.~Vereshagin and V.~Vereshagin, hep-th/0307256, submitted to
 {\em Phys. Rev. D}.
\bibitem{WeinQFT}
 S.~Weinberg,
 {\em The Quantum Theory of Fields}, 
 Cambridge University Press, Cambridge, 2000, vols. 1-3.
\bibitem{WeinEFT}
 S.~Weinberg,
 {\em Physica} {\bf 96A}, 327 (1979).
\bibitem{WeinQuasi}
 S.~Weinberg,
 {\em Phys. Rev.} {\bf 133}, 1B232 (1964).
\bibitem{Collins}
 J.~C.~Collins, {\em Renormalization},
 Cambridge University Press, Cambridge, 1984.
\bibitem{piNAVV}
 K.~Semenov-Tian-Shanski, A.~Vereshagin and V.~Vereshagin,
 ``Bootstrap and the parameters of pion-nucleon resonances'' in
 these Proceedings.
\bibitem{Shabat}
 B.~V.~Shabat, {\em An introduction to complex analysis},
 Nauka, Moscow, 1969 (in Russian).
\end{thebibliography}
\end{document}